\date{September 2018}
\pdfoutput=1

\documentclass[prl,twocolumn,preprintnumbers,longbibliography,amsmath,amssymb,superscriptaddress]{revtex4}

\usepackage{graphicx}
\graphicspath{{Figures/PNG files}{./}}
\usepackage{dcolumn}
\usepackage{bm}
\usepackage{float}
\usepackage{hyphenat}
\usepackage[dvipsnames]{xcolor}
\usepackage{physics}

\newcommand{\valencia}{Institute of Materials Science (ICMUV), University of Valencia, Catedr\'{a}tico Beltr\'{a}n 2, E-46980 Valencia, Spain}
\newcommand{\heriotwatt}{Institute of Photonics and Quantum Sciences, SUPA, Heriot-Watt University, Edinburgh EH14 4AS, UK}
\newcommand{\Tsukuba}{National Institute for Materials Science, Tsukuba, Japan}
\newcommand{\equalcontrib}{These authors contributed equally to this work}
\newcommand{\inl}{International Iberian Nanotechnology Laboratory (INL), Avda. Mestre Jos\'{e} Veiga s/n, 4715-330 Braga, Portugal}

\begin{document}

\title{Spin-layer locking of interlayer excitons trapped in moir\'e potentials}

\author{Mauro Brotons-Gisbert}
\thanks{\equalcontrib}
\email{M.Brotons_i_Gisbert@hw.ac.uk; h.baek@hw.ac.uk}
\affiliation{\heriotwatt}
\author{Hyeonjun Baek}
\thanks{\equalcontrib}
\email{M.Brotons_i_Gisbert@hw.ac.uk; h.baek@hw.ac.uk}
\affiliation{\heriotwatt}
\author{Alejandro Molina-S\'{a}nchez}
\affiliation{\valencia}
\affiliation{\inl}
\author{Aidan Campbell}
\affiliation{\heriotwatt}
\author{Eleanor Scerri}
\affiliation{\heriotwatt}
\author{Daniel White}
\affiliation{\heriotwatt}
\author{Kenji Watanabe}
\affiliation{\Tsukuba}
\author{Takashi Taniguchi}
\affiliation{\Tsukuba}
\author{Cristian Bonato}
\affiliation{\heriotwatt}
\author{Brian D. Gerardot}
\email{B.D.Gerardot@hw.ac.uk}
\affiliation{\heriotwatt}

\date{\today}

\begin{abstract}

\textbf{Van der Waals heterostructures offer attractive opportunities to design quantum materials. For instance, transition metal dichalcogenides (TMDs) possess three quantum degrees of freedom: spin, valley index, and layer index. Further, twisted TMD heterobilayers can form moir\'e patterns that modulate the electronic band structure according to atomic registry, leading to spatial confinement of interlayer exciton (IXs). Here we report the observation of spin-layer locking of IXs trapped in moir\'e potentials formed in a heterostructure of bilayer 2H-MoSe$_2$ and monolayer WSe$_2$. The phenomenon of locked electron spin and layer index leads to two quantum-confined IX species with distinct spin-layer-valley configurations. Furthermore, we observe that the atomic registries of the moir\'e trapping sites in the three layers are intrinsically locked together due to the 2H-type stacking characteristic of bilayer TMDs. These results identify the layer index as a useful degree of freedom to engineer tunable few-level quantum systems in two-dimensional heterostructures.}
\end{abstract}

\maketitle

The electronic and optical properties of van der Waals heterostructures can be widely engineered by the diverse choice of crystal combinations \cite{geim2013van} and their relative rotation \cite{bistritzer2011moire,zhang2017interlayer,cao2018unconventional,jin2019observation,alexeev2019resonantly} and interlayer spacing \cite{tong2017topological,yankowitz2018dynamic}.  For nearly commensurate bilayers (BLs), a slight lattice mismatch or relative rotation results in a moir\'e superlattice which spatially modulates the electronic band-structure. Single particle wavepackets can be trapped in the moir\'e-induced potential pockets with three-fold symmetry \cite{yu2017moire,wu2018theory,yu2018brightened}. Recently, signatures of IX trapped in such moir\'e  potentials were observed in TMD hetero-BL samples \cite{seyler2019signatures,tran2019evidence}.

\begin{figure*}
	\begin{center}
		\includegraphics[scale=0.95]{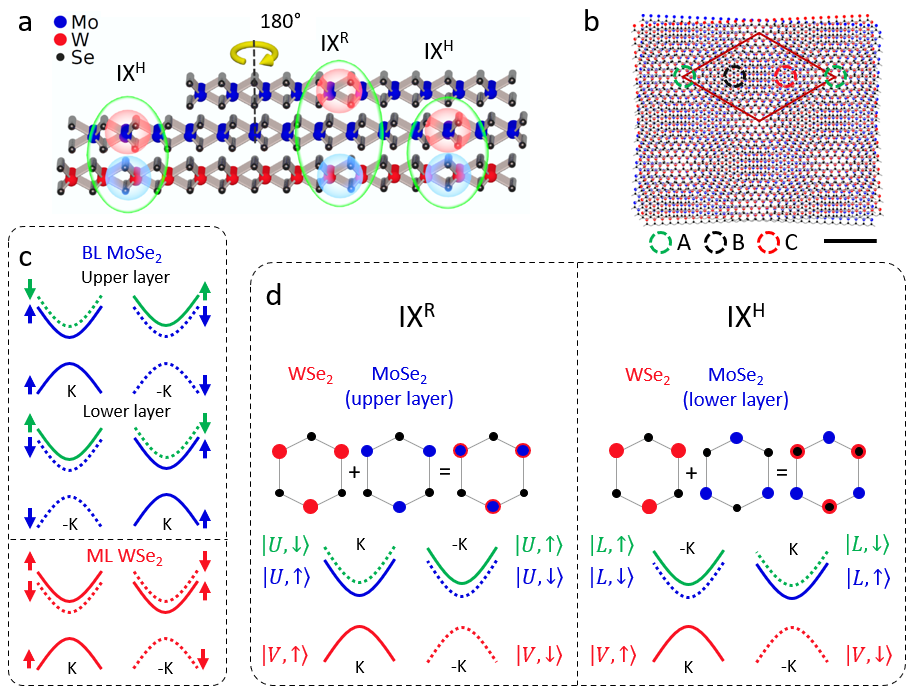}
	\end{center}
  \caption{\textbf{Spin-layer locking of moir\'e-trapped IXs in a van der Waals heterostructure.} \textbf{a}, Sketch of the heterostructures used in this work: a 2$H$-MoSe$_2$ crystal with ML- and BL-thick terraces stacked on top of a WSe$_2$ ML. The yellow arrow represents the 2$H$ stacking of BL MoSe$_2$. The red, blue and black dots represent the W, Mo, and Se atoms, respectively. The relative twists between the WSe$_2$ and the top and bottom MoSe$_2$ layers are $\sim 0^{\circ}$ (3$R$ stacking) and $\sim 60^{\circ}$ (2$H$ stacking), respectively. Electrons (red shadows) at the $\pm$K-valleys, localised either in the bottom or top layer MoSe$_2$, are strongly bound to holes (blue shadows) in the WSe$_2$, creating two species of IX (green circles): IX$^{H}$ and IX$^{R}$, respectively. \textbf{b}, Cartoon of the moir\'e superlattice formed in a $R$-stacked WSe$_2$/MoSe$_2$ heterostructure with a twist angle $\theta$ of 5$^\circ$. Scale bar, 2 nm. The red diamond represents a moir\'e supercell. The three highlighted regions represent moir\'e-trapping sites with different local atomic configurations: A ($R_h^h$), B ($R_h^X$) and C ($R_h^M$). \textbf{c}, Spin-valley locked configuration of BL 2$H$-MoSe$_2$ (top panel) and ML WSe$_2$ (bottom panel) at the $\pm$K corners of the hexagonal Brillouin zone. The red lines represent the electronic band structure at the direct band-gap edges of WSe$_2$. The blue lines denote the valence and conduction band-edges of MoSe$_2$, while the green lines represent the MoSe$_2$ CB excited state arising from spin-orbit coupling. The solid (dotted) lines depict spin-up (down) band-edge states. Up (down) arrows indicate spin-up (down) conduction-band and valence-band electrons. \textbf{d}, Valley- and spin-layer pairing of the electrons for IX$^{R}$ (left) and IX$^{H}$  (right)  together with the top views of the corresponding stacking configurations. The labels $\ket{V,\uparrow/\downarrow}$, $\ket{L,\uparrow/\downarrow}$, and $\ket{U,\uparrow/\downarrow}$ represent the band-edge states at $\pm$K, with $V$ the WSe$_2$ valence band, $U$ ($L$) the upper (lower) MoSe$_2$ layer (layer index), and $\uparrow$ ($\downarrow$) the electron spin $s_z$ = 1/2 (-1/2). The red, blue and black dots represent the atoms in the WSe$_2$ and MoSe$_2$ layers according to \textbf{a}.}
  \label{fig1}
\end{figure*}

Due to the potential to harness the carrier spin, valley index, and layer index \cite{xu2014spin}, layered TMDs present an intriguing platform for quantum electronics and optics. The broken inversion symmetry in TMDs leads to an effective coupling between a carrier spin and the valley index of the electrons or holes at the $\pm$K corners of the hexagonal Brillouin zone \cite{xiao2012coupled}, as shown for monolayer (ML) WSe$_2$ in Fig. \ref{fig1}c. Like the real spin, the valley index is associated to a magnetic moment \cite{xiao2012coupled, xu2014spin}, which results in valley dependent selection rules. Light with $\sigma^{\pm}$-polarisation creates electron-hole pairs exclusively in the $\pm$K valley, enabling optical and magnetic manipulation of intralayer valley excitons \cite{mak2012control, zeng2012valley, aivazian2015magnetic, srivastava2015valley}. Layer index is found in 2$H$-type BL TMDs which have a $180^\circ$ in-plane rotation between the top- and bottom-layer and minimal interlayer electronic hopping at the K valleys \cite{xiao2012coupled, gong2013magnetoelectric, jones2014spin, riley2014direct}. The in-plane rotation inverts the valley alignment in the two layers, such that a unique spin-valley configuration is locked to each layer, as shown for BL MoSe$_2$ in Fig. \ref{fig1}c. This gives rise to the layer index. Additionally, the atomic registries of the two layers are intrinsically locked together in 2$H$-type BL TMDs.

Owing to these quantum degrees of freedom, TMDs are ideal ingredients to realise the concept of exciton trapping in moir\'e potentials \cite{yu2017moire, wu2018theory, yu2018brightened, seyler2019signatures, tran2019evidence}. Two different ML TMDs can be combined with an atomically sharp interface to create a hetero-BL system with Type II-band alignment \cite{chiu2015determination, wilson2017determination}, which favours spatial separation of photogenerated carriers: electrons and holes rapidly transfer to reside in the layer with the lowest energy conduction (CB) and valence band-edge (VB), respectively \cite{hong2014ultrafast}. Due to strong Coulomb interaction, the electrons and holes form spatially-indirect interlayer excitons which exhibit the valley-dependent optical selection rules of the monolayer TMDs but with longer exciton lifetimes, robust spin-valley polarisation, and large electric field tunability \cite{yu2015anomalous, rivera2015observation, rivera2016valley, hanbicki2018double, ciarrocchi2019polarization}. Hetero-BLs with arbitrary stacking angles can be fabricated by transfer of mechanically exfoliated flakes \cite{seyler2019signatures, tran2019evidence, wilson2017determination, rivera2015observation, ciarrocchi2019polarization} or chemical vapour deposition \cite{zhang2017interlayer, chiu2015determination, hong2014ultrafast, hanbicki2018double}. Nearly aligned $0^\circ$ (\textit{R}-type) or $60^\circ$ (\textit{H}-type) stacking of the hetero-BL yields minimum displacement in momentum space for carriers at the band edges ($\pm$K), maximizing coupling to the light cone for optical transitions \cite{yu2015anomalous}. As shown in Fig. \ref{fig1}d, IXs in an \textit{R}-type stacked hetero-BL (IX$^{R}$) have valley conserving ground state optical transitions, identical to monolayer TMDs. Conversely, IXs in \textit{H}-type stacked materials (IX$^{H}$) have valley contrasting optical transitions. We label the band-edge states at $\pm$K as $\ket{V,\uparrow/\downarrow}$, $\ket{L,\uparrow/\downarrow}$, and $\ket{U,\uparrow/\downarrow}$, where $V$ represents the WSe$_2$ valence band, $U$ ($L$) is the upper (lower) MoSe$_2$ layer (layer index), and $\uparrow$ ($\downarrow$) represents the electron spin $s_z$ = 1/2 (-1/2). In addition, nearly aligned $0^\circ$ or $60^\circ$ stacking yields a moir\'e potential landscape in a hetero-BL in which three trapping sites (A, B, and C) with atomic registries $R_h^h$ (A), $R_h^X$ (B) and $R_h^M$ (C) emerge (see Fig. \ref{fig1}b), where $R^\mu_h$ denotes an $R$-type stacking with the $\mu$ site of the electron layer (either $h$ the hexagon centre, $X$ the chalcogen site or $M$ the metal site) vertically aligned with the hexagon centre ($h$) of the hole layer \cite{yu2017moire, tong2017topological, yu2018brightened}.

Here we explore the spin, valley index, and layer index properties of moir\'e-trapped IXs in an artificial trilayer (TL) heterostructure consisting of a ML WSe$_2$ and BL 2$H$-MoSe$_2$ (see Fig. \ref{fig1}a). Due to the $d_{z^2}$ symmetry of the orbitals at the conduction band edge, electrons in BL MoSe$_2$ present a vanishing interlayer hopping at the $\pm$K points, which leads to a strong coupling between the electron spin, the layer and the valley degrees of freedom \cite{gong2013magnetoelectric}. We exploit the spin-layer locking phenomenon of BL 2$H$-MoSe$_2$ to probe two moir\'e-trapped IX species with contrasting spin-valley alignment: the holes, localised in the WSe$_2$ monolayer, are strongly Coulomb bound to electrons localised in either the lower or upper MoSe$_2$ layer to form IX$^{H}$ or IX$^{R}$ species, respectively. Each localised IX species has a distinct spin-layer-valley configuration: IX$^{R}$ (IX$^{H}$) exhibit electron-hole pairs with parallel (antiparallel) spin-valley-locked magnetic moments. The emission from each IX species exhibits circular polarisation which, when combined with knowledge of the spin and valley configuration, enable determination of the atomic registry of the moir\'e trapping sites. We report the observation of a new moir\'e-trapping site for IX$^{R}$ excitons in the TL-heterostructure as compared to IX$^{R}$ excitons in WSe$_2$/MoSe$_2$ hetero-BLs \cite{seyler2019signatures}, which we attribute to the 2$H$-type stacking characteristic of BL MoSe$_2$ that results in the intrinsic locking of the atomic registries of the three layers. Our results present new evidence that add confidence to the moir\'e potential as the origin of the IX confinement.

\section{Trapped interlayer excitons in a trilayer heterostructure}

\begin{figure*}[t]
	\begin{center}
		\includegraphics[scale=0.54]{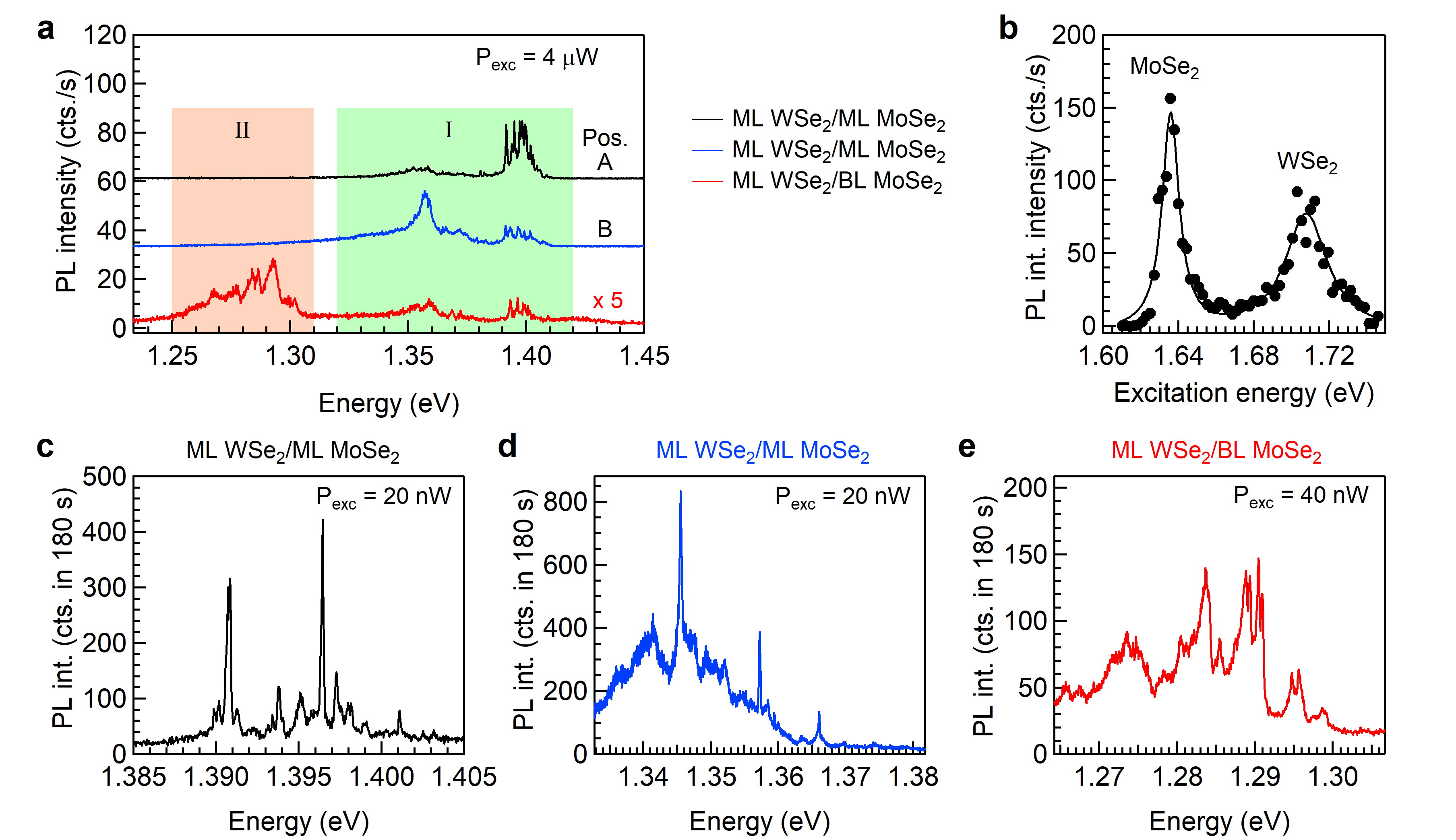}
	\end{center}
  \caption{\textbf{Optical spectroscopy properties of spin-layer-locked IXs trapped in moir\'e potentials.} \textbf{a}, Low-temperature confocal PL spectra measured at three spatial positions of the heterostructure. The black and blue positions (A and B, respectively) represent spectra corresponding to different positions of the ML MoSe$_2$/ML WSe$_2$ heterostructure, with emission in energy range I: 1.32 - 1.42 eV (green shaded area). The red line corresponds to a spectrum measured in the ML WSe$_2$/BL MoSe$_2$ region of the heterostructure, with emission in energy range II: 1.25 - 1.31 eV (red shaded area). \textbf{b}, Photoluminescence excitation intensity plot of a representative IX (black dots), showing two resonances corresponding to the intralayer exciton states in ML MoSe$_2$ and ML WSe$_2$. The black solid line represents a fit of the experimental data to two Lorentzian peaks. \textbf{c-e}, Optical spectra of representative IXs trapped in moir\'e potentials in the ML WSe$_2$/ML MoSe$_2$ (\textbf{c} and \textbf{d}), and the ML WSe$_2$/BL MoSe$_2$ regions of the heterostructure (\textbf{e}) under low excitation power.}
  \label{fig2}
\end{figure*}

To justify our choice of TL heterostructure for the realisation of spin-layer locked IXs, we performed \textit{ab initio} calculations (see Suppl. Note 1). The artificial TL heterostructures consisted of a 2$H$-MoSe$_2$ crystal with ML and BL terraces mechanically stacked on top of a WSe$_2$ ML and encapsulated by hBN in an inert environment (see Suppl. Note 2). The 2$H$ stacking of ML WSe$_2$ and the bottom ML MoSe$_2$ was confirmed by linear-polarisation-resolved second harmonic generation measurements (see Suppl Note 3). Figure \ref{fig2}a shows representative low-temperature (T = 4 K) confocal photoluminescence (PL) spectra measured using continuous wave excitation at 2.33 eV and an excitation power ($P_{exc}$) of 4 $\mu$W. PL spectra corresponding to different positions of the ML MoSe$_2$/ML WSe$_2$ heterostructure (black and blue spectra) show emission in the energy range 1.32 - 1.42 eV (energy range I). We observe that the emission is centred around two main energy windows (1.385 - 1.405 eV and 1.345 - 1.375 eV), but with spatially dependent relative intensities. Positions A (black spectrum) and B (blue spectrum) correspond to the spatial positions with the brightest relative intensities for the two windows. To confirm that the observed emission arises from IXs, we performed PL excitation spectroscopy, scanning a continuous-wave excitation laser from 1.61 eV to 1.75 eV while monitoring the intensity of the emission peaks. Figure \ref{fig2}b shows a representative PL excitation spectrum, featuring two prominent resonances which correspond to the absorption of the intralayer 1$s$ exciton states in ML MoSe$_2$ and WSe$_2$. PL spectra in the TL part of the heterostructure is markedly different: emission in spectral region I is still observed but with reduced intensity relative to a new band of emission at lower energy (1.25 - 1.31 eV, region II), see for example the red spectrum in Fig. \ref{fig2}a. The spatial dependence of the PL spectra suggests that ranges I and II originate from different IX species, as expected from the stacking configuration presented in Fig. \ref{fig1}. This heterostructure stacking configuration is corroborated by means of DC Stark effect measurements (see Suppl. Note 4). Therefore, we label IXs with emission in ranges I and II as IX$^H$ and IX$^R$, respectively. The PL emission energy of IX$^H$ matches well with recent reports for IX emission in MoSe$_2$/WSe$_2$ hetero-BLs \cite{rivera2015observation, hanbicki2018double, ciarrocchi2019polarization, seyler2019signatures, Torun2018}, while IX$^R$ emission in TL heterostructures have yet to be explored in such detail. 

With reducing $P_{exc}$, the broad PL gradually disappears until the sharp peaks dominate, as shown in Suppl. Note 5. Here, high-resolution PL spectra measured with reduced $P_{exc}$ reveal that IX$^{H}$ (IX$^{R}$) PL peaks exhibit Lorentzian lineshapes with average full-width at half maximum (FWHM) of $\sim$100 $\mu$eV ($\sim$250 $\mu$eV); the narrowest peaks observed are $\sim$70 $\mu$eV (see Suppl. Note 5). These linewidths, two orders of magnitude narrower than typical IX linewidths in WSe$_2$/MoSe$_2$ hetero-BLs \cite{rivera2016valley, jiang2018microsecond, hanbicki2018double, ciarrocchi2019polarization}, are comparable to quantum emitters in ML WSe$_2$ \cite{srivastava2015optically, tonndorf2015single, brotons2019coulomb} and to recently reported moir\'{e}-trapped excitons  \cite{seyler2019signatures}. Supplementary Figure 7 shows the full evolution of the IX$^{H}$ emission under increasing $P_{exc}$. For each species of IX, the intensity saturates with increasing $P_{exc}$ according to a two-level system (Suppl. Fig. 7). Additionally, we observe minimal spectral wandering ($<$ 10 $\mu$eV) at long time scales (see Suppl. Fig. 8).  

\begin{figure*}[t]
	\begin{center}
		\includegraphics[scale=0.6]{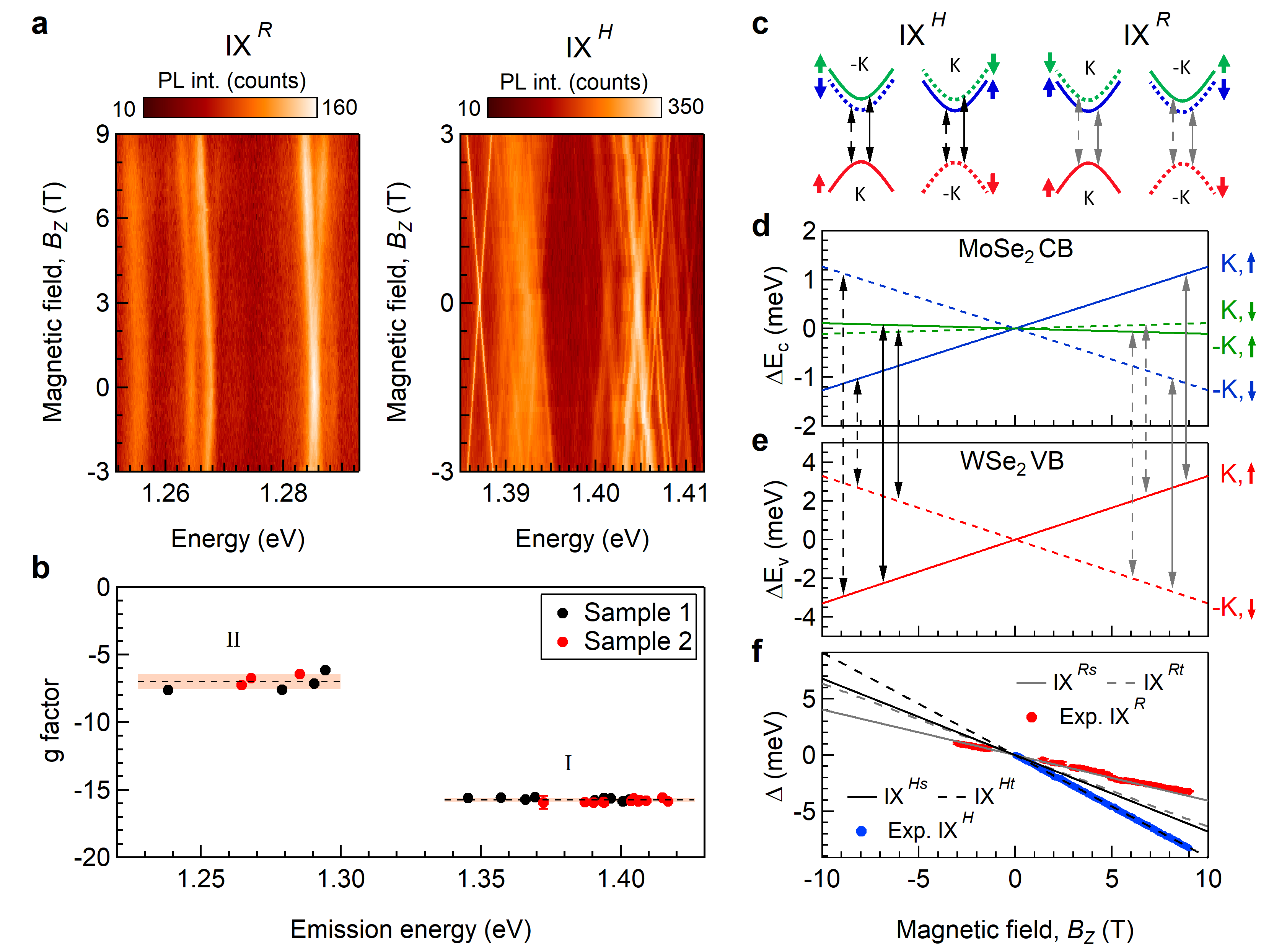}
	\end{center}
  \caption{\textbf{Magneto-optic properties of spin-layer-locked IXs trapped in moir\'e potentials.} \textbf{a}, Magnetic field dependence of representative moir\'e-trapped IX$^R$ (left) and IX$^H$ excitons (right) as a function of an applied out-of-plane magnetic field. \textbf{b}, Summary of the $g$-factor values measured for moir\'e-trapped IX$^R$ and IX$^H$ excitons in heterostructures 1 (black dots) and 2 (red dots) as a function of their emission energies. The horizontal dashed lines and the shaded areas represent the average value and standard deviation of the $g$-factors, respectively. \textbf{c}, Schematics of the spin-valley configuration for IX$^R$ (left) and IX$^H$ (right) excitons. The vertical arrows represent the optical transitions for each IX exciton and spin configuration. \textbf{d, e}, Zeeman shifts for the conduction ($\Delta$E$_C$) and valence ($\Delta$E$_V$) band edges calculated by using Eqs. (\ref{Eq:Zeeman_AAs})-(\ref{Eq:Zeeman_ABt}). The labels K (-K) and $\uparrow$ ($\downarrow$) represent the valley index and spin configuration of the corresponding band edges, respectively. Grey (black) arrows represent optical transitions in IX$^{R}$ (IX$^{H}$) excitons, while solid (dashed) lines represent spin-singlet (spin-triplet) optical transitions.  \textbf{f}, $B_z$-dependence of the $\Delta$ value for  for each IX exciton species and spin configuration as calculated from Eqs. (\ref{Eq:Zeeman_AAs})-(\ref{Eq:Zeeman_ABt}). IX$^{Hs}$, IX$^{Rs}$, IX$^{Ht}$ and IX$^{Rt}$, represent IXs arising from spin-singlet ($s$) and spin-triplet ($t$)  optical  transitions, respectively. The colour and line style used for the $\Delta$ value of each IX configuration is consistent the optical transitions of the corresponding excitons in \textbf{d} and \textbf{e}. Red and blue dots represent the $B_z$-dependence of the experimental $\Delta$ values of representative trapped IX$^{R}$ and IX$^{H}$ excitons, respectively, as extracted from Lorentzian fits of the experimental data. Here the error bar size is smaller than the data points.} 
  \label{fig3}
\end{figure*}

\section{Spin-valley-layer configurations of trapped interlayer excitons}

The narrow linewidths and saturation behaviour provide strong evidence of IXs trapped in a moir\'e confinement potential. For the low $P_{exc}$ used in the PL spectra shown in Figs. \ref{fig2}c-e, the IX density is not high enough to fill all the trapping sites of the moir\'e lattice within the diffraction limited focus of our confocal PL spot.  Instead, we only observe emission from a few trapping sites (10 to 20 depending on the spatial position in the sample). Under increasing excitation power, as more trapping sites are filled, we progressively lose the ability to resolve individual spectral lines.

To confirm these peaks arise from band-edge states and disentangle the spin-layer-valley configuration of each exciton species, we perform magneto-optical spectroscopy measurements in Faraday configuration. Figure \ref{fig3}a shows the magnetic field ($B_z$) dependence of representative moir\'e-trapped IX$^{R}$ and IX$^{H}$ (left and right panel, respectively). A clear linear Zeeman splitting with increasing $B_z$ is observed for every peak, and it is immediately noticeable that the $g$-factor of IX$^{H}$ excitons is considerably larger than IX$^{R}$ excitons. To confirm these are universal features of moir\'e trapped IX$^{H}$ and IX$^{R}$, a second WSe$_2$/MoSe$_2$ heterostructure with both ML WSe$_2$/ML MoSe$_2$ and ML WSe$_2$/BL MoSe$_2$ regions and similar stacking angle ($\sim60^{\circ}$ between the ML WSe$_2$/BL MoSe$_2$) was fabricated and measured (Sample 2). Sample 2 also exhibits IX$^{R}$ and IX$^{H}$ with narrow emission peaks in the same energy ranges as Sample 1. Figure \ref{fig3}b summarises the measured $g$-values for each IX species in Samples 1 and 2 (black and red dots, respectively) as a function of their emission energy. Trapped IX$^{R}$, which only appear in energy range II from the TL part of the samples, exhibit a $g$-factor of -7.0 $\pm$ 0.6. On the other hand, trapped IX$^{H}$ in energy range I are observed both in the hetero-BL and hetero-TL regions and exhibit a $g$-factor of -15.76 $\pm$ 0.13. 

The striking $B_z$ dependence for each IX species has its origin in their corresponding spin-valley configuration \cite{seyler2019signatures} (see Fig. \ref{fig3}c), demonstrating that IX$^{R}$ and IX$^{H}$ preserve the spin-layer locking of electrons in BL 2$H$-MoSe$_2$. Since both carrier spin and valley index are associated to a magnetic moment \cite{xiao2012coupled, xu2014spin}, the total valley-selective splitting of the interlayer transitions ($\Delta$) amounts to \cite{nagler2017giant, seyler2019signatures}
\begin{align}
    \Delta\left(B_z\right) = E^{\sigma^+}\left(B_z\right)-E^{\sigma^-}\left(B_z\right) = g\mu_BB_z,
    \label{Eq:Zeeman_split}
\end{align}
with $g$ being an effective $g$-factor, $\mu_B$ being the Bohr magneton, and $E^{\sigma^\pm}\left(B_z\right)$ being the $B_z$-dependent energies of the intervalley transitions with $\sigma^\pm$ polarisation. $E^{\sigma^\pm}$ is defined as the energy difference between the conduction ($E_c^\pm$) and valence ($ E_v^\pm$) band edges associated to $\sigma^\pm$ transitions ($E^{\sigma^\pm}\left(B_z\right) = E_c^\pm - E_v^\pm$). The Zeeman shift of the $E_c^\pm$ ($\Delta E_c$) and $E_v^\pm$ ($\Delta E_v$) band edges can be estimated as a combination of three different magnetic moment contributions: $\Delta E_{c/v} = \Delta_{s}+\Delta_{a}+\Delta_{v}$, where $\Delta_s = 2s_z\mu_BB_z$ is the spin contribution, $\Delta_a = l_{c/v}\mu_BB_z$ is the atomic orbital contribution, and $\Delta_v = \tau \alpha_{c/v}\mu_BB_z$ represents the valley contribution arising from the Berry curvature \cite{xiao2012coupled}. Here, $s_z = \pm 1/2$ is the electron spin, $l_{c} = 0$ ($l_{v} = 2\tau$) is the magnetic quantum number for the atomic orbital at the conduction (valence) band edge, $\tau = \pm1$ is the index for the $\pm$K valleys, and $\alpha_c$ ($\alpha_v$) is the magnetic moment of the conduction (valence) band edge \cite{aivazian2015magnetic}. According to the leading order of a simplified \textit{\textbf{k$\cdot$p}} approximation for the band-edge carriers \cite{xiao2012coupled, yao2008valley}, the valley magnetic moments of the conduction and valence band can be estimated as $\alpha_{c,v} = m_0/m^*_{e,h}$, with $m_{0}$ the free electron mass and $m^*_{e}$ ($m^*_{h}$) the electron (hole) effective mass at the conduction (valence) band edge. Figures \ref{fig3}d and \ref{fig3}e show the Zeeman shifts of the conduction (MoSe$_2$) and valence (WSe$_2$) band edges of a WSe$_2$/MoSe$_2$ heterostructure, respectively, calculated using $m^*_{h} = 0.37$ $m_{0}$ \cite{kormanyos2015k} and $m^*_{e} = 0.84$ $m_{0}$ \cite{larentis2018large, goryca2019revealing}. Here we assume that the spin-split conduction bands at the $\pm$K points of ML and BL MoSe$_2$ have similar effective masses \cite{larentis2018large}, and therefore we use the same $\alpha_{c}$ for both. The colour and line style used for each band edge are consistent with the schematics of the spin-layer configuration shown in Fig. \ref{fig3}c. Solid (dashed) lines represent the Zeeman shifts of band edges with spin up (down). Red lines represent the Zeeman shifts of the valence band states in WSe$_2$, whereas blue and green lines represent the Zeeman shifts of the MoSe$_2$ conduction band edges with parallel and antiparallel valley and spin configuration, respectively. 

According to Equation (\ref{Eq:Zeeman_split}), the different spin-layer-valley configuration of IX$^{R}$ and IX$^{H}$ excitons results in different total Zeeman splittings $\Delta$. Therefore, the $g$-factor of the moir\'e-trapped valley excitons is representative of their spin-valley configuration, which, as a consequence of the spin-layer locking of electrons in BL MoSe$_2$, also indicates whether the electron is localised in the bottom or top MoSe$_2$ layer. Furthermore, two different $\Delta$  splittings are possible for both IX$^{R}$ and IX$^{H}$ excitons depending on the conduction bands involved in the optical transitions (see Methods). 

The vertical arrows in Figs. \ref{fig3}d and \ref{fig3}e represent the optical transitions responsible of the Zeeman splittings for each IX exciton and spin configuration. Grey (black) arrows represent optical transitions in IX$^{R}$ (IX$^{H}$) excitons, while solid (dashed) lines represent spin-singlet (spin-triplet) optical transitions. Figure \ref{fig3}f shows the $B_z$-dependence of the $\Delta$ value  for each IX exciton and spin configuration as calculated from Equations (\ref{Eq:Zeeman_AAs})-(\ref{Eq:Zeeman_ABt}). The colour and line style used for the $\Delta$ value of each IX configuration is consistent with the one employed to indicate the optical transitions of the corresponding excitons. Figure \ref{fig3}f also shows the $B_z$-dependence of the experimental $\Delta$ values of two representative trapped IX$^{R}$ (red dots) and IX$^{H}$ (blue dots) excitons, as extracted from Lorentzian fits of the experimental data. The good agreement observed between the calculated and experimental $\Delta$ values corroborates our initial identification of IX$^{R}$ and IX$^{H}$ excitons. Moreover, the magneto-optical measurements provide additional information. The results shown in Fig. \ref{fig3}f indicate that the observed IX$^{R}$ and IX$^{H}$ excitons arise from optical transitions involving the lowest spin-split conduction band of MoSe$_2$ at $\pm$K. This observation leads to spin-conserved and spin-flip optical transitions for IX$^{R}$ and IX$^{H}$ excitons, respectively. Although the latter is normally forbidden in ML TMDs due to its spin-flip nature, it can be brightened due to the selection rules dictated by the resulting interlayer atomic registry of the moir\'e pattern in our heterostrtuctures \cite{yu2018brightened}.

The spin-layer-locked nature of IX$^{R}$ and IX$^{H}$ allows us to estimate the magnetic moment contribution of the conduction and valence band edges, and therefore the corresponding carrier effective masses (see Methods). 

\begin{figure*}[t]
	\begin{center}
		\includegraphics[scale=0.6]{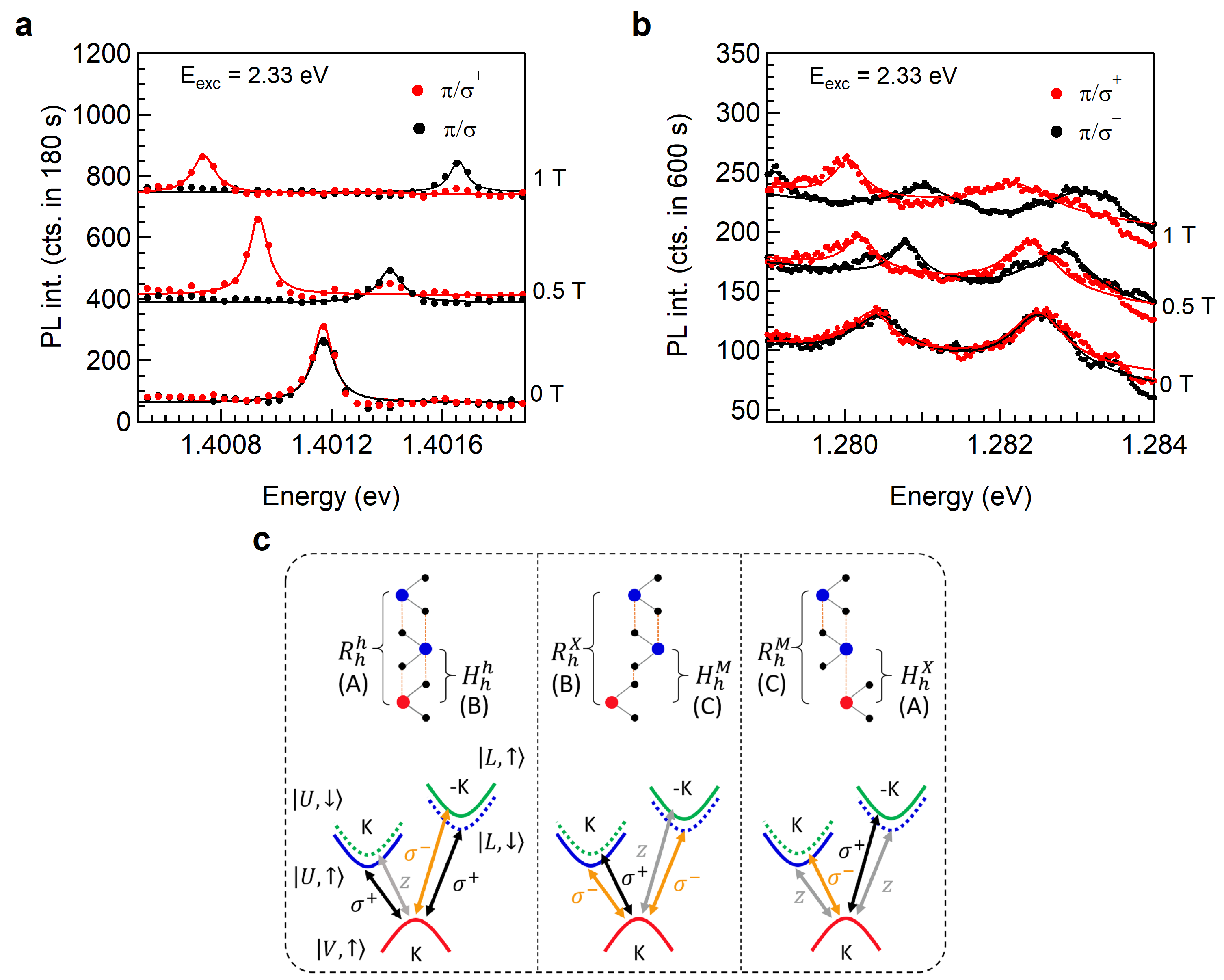}
	\end{center}
  \caption{\textbf{Optical selection rules of spin-layer-locked IXs trapped in moir\'e potentials.} \textbf{a-b}, PL spectra of representative IX$^{H}$ and IX$^{R}$ excitons, respectively, under linearly polarised excitation at 2.33 eV and different applied magnetic fields for circularly-polarised-resolved $\sigma^+$ (red dots) and $\sigma^-$ (back dots) collection. The red and black solid lines represent Lorentzian fits to the experimental data. \textbf{c}, Schematics of the selection rules for optical transitions involving the K-point valence band for both spin-singlet and spin-triplet IXs trapped in moir\'e potential sites with different atomic registries \cite{yu2018brightened}. Black, orange and grey double arrows represent optical transitions with $\sigma^+$, $\sigma^-$, and $z$ polarisation. $R(H)^\mu_h$ denotes an $R (H)$-type stacking with the $\mu$ site of the electron layer (either $h$ the hexagon centre, $X$ the chalcogen site or $M$ the metal site) vertically aligned with the hexagon centre ($h$) of the hole layer.}
  \label{fig4}
\end{figure*}

\section{Local atomic registries of moir\'e trapping sites}

The relative Zeeman shifts of the conduction and valence band edges at $\pm$K points enable further insight, based on the fact that the $B_z$-induced shift of the WSe$_2$ valence band edge is larger than the corresponding shifts of the MoSe$_2$ conduction bands (as shown in Figs. \ref{fig3}d and \ref{fig3}e). The smaller Zeeman splitting for the conduction band edges is a consequence of two factors: the smaller magnetic moment contribution from the atomic orbitals of the conduction band ($l_{c} = 0$), and the higher effective mass of electrons in the conduction band of MoSe$_2$. The larger Zeeman shift for the valence band edges leads to a striking consequence: for $B_z>0$, the energy of the optical transitions involving the valence band edges of WSe$_2$ at K (-K) always shift to lower (higher) energies regardless of the spin-valley configuration of the IX. Based on this, helicity-resolved PL measurements can provide information about the nature of the moir\'e confinement potential. Figures \ref{fig4}a and \ref{fig4}b show circularly-polarised-resolved spectra of representative IX$^{H}$ and IX$^{R}$, respectively, under linearly polarised ($\pi$) excitation at 2.33 eV and different applied magnetic fields. Both IX$^{H}$ and IX$^{R}$ exhibit strong circular polarisation, and application of $B_z$ results in a Zeeman shift of the $\sigma^+$-polarised ($\sigma^-$-polarised) PL peak towards lower (higher) energies. For zero magnetic field, energy-degenerate $\sigma^+$ and $\sigma^-$ emissions are observed with near identical intensity, indicating that the IXs are trapped in confinement potentials that preserve the rotational $C_3$ symmetry. The degenerate circularly-polarised emission of IX$^{R}$ and IX$^{H}$ contrasts with the emission polarisation properties of neutral excitons in quantum emitters in ML WSe$_2$, which typically exhibit a large fine-structure splitting and strictly linear polarisation \cite{srivastava2015optically, tonndorf2015single, brotons2019coulomb} arising from the electron-hole exchange interaction energy and asymmetry in the confinement potential. The absence of observable fine-structure splitting is only expected in cases for which rotational symmetry of the crystal lattice and confining potential is maintained.

The polarisation selection rules of the moir\'e-trapped excitons are dictated by the local atomic registry of the moir\'e trapping site \cite{yu2017moire,yu2018brightened}. Figure \ref{fig4}c shows the selection rules for optical transitions involving the K-point valence band for both spin-singlet and spin-triplet IXs trapped in moir\'e potential sites with different atomic registries \cite{yu2018brightened}. Accordingly, the results in Figs. \ref{fig4}a and \ref{fig4}b indicate that IX$^{R}$ and IX$^{H}$ originate from IXs trapped in moir\'e potentials with interlayer atomic registries $R^h_h$ (A) and $H^h_h$ (B), respectively. Interestingly, we find that IX$^{R}$ excitons in the TL heterostructure present a comparable $g$-factor magnitude but opposite valley selection rules compared to IX$^{R}$ excitons in WSe$_2$/MoSe$_2$ hetero-BLs \cite{seyler2019signatures}. These results indicate that localised IX$^{R}$ excitons in the TL heterostructure and in WSe$_2$/MoSe$_2$ hetero-BLs \cite{seyler2019signatures} are the same IX species (same spin-valley pairing) trapped in moir\'e potentials with different atomic registries. We attribute the different atomic registry of the moir\'e trapping potentials for IX$^{R}$ excitons in the TL heterostructure ($R^h_h$) and the WSe$_2$/MoSe$_2$ hetero-BLs ($R^X_h$) to the 2$H$-type stacking of characteristic of BL MoSe$_2$, which results in the intrinsic locking of the atomic registries of the three layers (see left panel of Fig. \ref{fig4}c).

\section{Discussion and outlook}

The magneto-optical spectroscopy experiments reveal the remarkable consequence of spin-layer locking of moir\'e-trapped excitons in the artificial TL heterostructures. Two IX species, IX$^{R}$ and IX$^{H}$, with distinct spin-valley-layer configurations are observed: IX$^{R}$ (IX$^{H}$) composed of electrons in the top (bottom) MoSe$_2$ layer present carriers with parallel (antiparallel) spin-valley locked contributions, resulting in an effective layer-locking of the Land\'e g-factors of the trapped IX. At cryogenic temperatures, both trapped IX species exhibit narrow linewidths and saturate with increasing excitation power, hallmarks of a few-level quantum confined system. An unambiguous demonstration of quantum emission from the moir\'e-trapped excitons, for instance photon antibunching, remains an important target. Finally, the combination of magneto-optics and helicity-resolved PL measurements allows the determination of the atomic registries of the moir\'e trapping sites. 

For quantum information applications, a basic requirement is the ability to initialise the state of the few-level quantum system, which has been achieved for valley qubits in ML TMDs via optical pumping \cite{mak2012control, zeng2012valley}. Here we show the trapped IXs retain the strong valley polarisation of the constituent ML semiconductors (see Suppl. Fig. 10). 

Finally, we remark that the phenomenon of spin-layer locking provides a means to engineer few-level quantum systems in van der Waals heterostructures. For example, the selection rules arising from the $C_3$ symmetry for the atomic registries we observe (left panel of of Fig. \ref{fig4}c), create a ``vee-type" three level system: a single ground state ($\ket{V,\uparrow}$) couples to two non-degenerate excited states ($\ket{L,\downarrow}$ and $\ket{U,\uparrow}$). Alternatively, a three-level system analogous to a ``spin-lambda" type atom, in which two ground states couple to a common excited state, can be engineered using the layer-index degree of freedom in a TL heterostructure consisting of BL 2$H$-WSe$_2$/ML MoSe$_2$. This quantum state engineering opens new opportunities for advanced quantum control techniques in the van der Waals platform.


\section{Acknowledgements}
The authors thank J. J. Finley, K. M\"uller, M. Kremser, and A. H\"{o}gele for discussions. This work is supported by the EPSRC (grant no. EP/P029892/1 and EP/S000550/1), the ERC (grant no.  725920) and the EU Horizon 2020 research and innovation program under grant agreement no. 820423. AMS acknowledges the Juan de la Cierva (Grant IJCI-2015-25799) program (MINECO, Spain) and the Marie-Curie-COFUND program Nano TRAIN For Growth II (Grant Agreement 713640). The computations were performed on the Tirant III cluster of the Servei d'Inform\`{a}tica of the University of Valencia (project vlc82). Growth of hBN crystals by K.W. and T.T. was supported by the Elemental Strategy Initiative conducted by the MEXT, Japan and the CREST (JPMJCR15F3).
B.D.G. is supported by a Wolfson Merit Award from the Royal Society and a Chair in Emerging Technology from the Royal Academy of Engineering.

\section{Author contributions}
B.D.G. conceived and supervised the project. H.B. fabricated the samples. K.W. and T.T. supplied the hBN crystals. M.B.-G. and H.B. performed the experiments, assisted by A.C., C.B. and D.W. M.B.-G. analysed the data, assisted by E.S. and B.D.G. A.M.-S. performed the ab initio calculations. M.B.-G. and B.D.G. cowrote the paper with input from all authors. M.B.-G. and H.B. contributed equally to this work.

\section{Competing interests}
The authors declare no competing interests.

\newpage
\clearpage

\section{Methods}

\subsection{Zeeman splitting for IXs with different  spin-layer-valley  configurations}

\begin{small}
Optical transitions between the bottom conduction band states and the top valence band state at K give rise to four possible IX configurations: IX$^{Hs}$ ($\ket{V,\uparrow}\leftrightarrow\ket{L,\uparrow}$), IX$^{Ht}$ ($\ket{V,\uparrow}\leftrightarrow\ket{L,\downarrow}$), IX$^{Rs}$ ($\ket{V,\uparrow}\leftrightarrow\ket{U,\uparrow}$), and IX$^{Rt}$ ($\ket{V,\uparrow}\leftrightarrow\ket{U,\downarrow}$), where $s$ and $t$ superscripts  denote  spin-conserving (spin-singlet)  and  spin-flip  (spin-triplet)  optical  transitions, respectively.\\

Using Eq. (\ref{Eq:Zeeman_split}), IX$^{R}$ excitons can take $\Delta$ values of:
\begin{align}
    \lvert \Delta E^{Rs}\left(B_z\right)\rvert = 2 \lvert 2+\alpha_v - \alpha_c\rvert \mu_BB_z,
    \label{Eq:Zeeman_AAs}
\end{align}
and
\begin{align}
    \lvert \Delta E^{Rt}\left(B_z\right)\rvert = 2 \lvert 4+\alpha_v - \alpha_c\rvert \mu_BB_z;
    \label{Eq:Zeeman_AAt}
\end{align}
while IX$^{H}$ excitons can present $\Delta$ values of:
\begin{align}
    \lvert \Delta E^{Hs}\left(B_z\right)\rvert = 2 \lvert 2+\alpha_v + \alpha_c\rvert \mu_BB_z,
    \label{Eq:Zeeman_ABs}
\end{align}
and
\begin{align}
    \lvert \Delta E^{Ht}\left(B_z\right)\rvert = 2 \lvert 4+\alpha_v + \alpha_c\rvert \mu_BB_z.
    \label{Eq:Zeeman_ABt}
\end{align}
\end{small}

\subsection{Magnetic moment contributions of the band edges}
\begin{small}
From the combination of Equations (\ref{Eq:Zeeman_AAs}) and (\ref{Eq:Zeeman_ABt}), and the measured $g$-factors for IX$^{H}$ and IX$^{R}$, we estimate magnetic moment contributions of $\alpha_v$ = 2.69 $\pm$ 0.15 and $\alpha_c$ = 1.19 $\pm$ 0.15. The estimated magnetic moment for the band edges yield effective masses of $m^*_{e} \approx 0.84 \pm 0.11$ $m_0$ and $m^*_{h} \approx 0.37 \pm 0.02$ $m_0$ for electrons and holes at the bottom conduction band of MoSe$_2$ and top valence band of WSe$_2$, respectively.
\end{small}

\begin{small}
In the previous discussion we have considered that IX$^{H}$ arise from optical spin-flip transitions involving the lowest spin-split conduction band of MoSe$_2$ at $\pm$K. However, IX$^{H}$ could also be tentatively attributed to spin-conserved optical transitions from the top valence band of WSe$_2$ and the higher energy spin-split conduction band of MoSe$_2$ at $\pm$K \cite{seyler2019signatures} (see Fig. 3c). For such spin-singlet configuration (IX$^{Hs}$), the spin magnetic moments contribution of the electron and hole cancel each other, and the total Zeeman splitting can be calculated by Eq. (4). In this case, the combination of the measured $g$-factors with Eqs. (2) and (4) leads to effective masses of $m^*_{e} \approx 0.46 \pm 0.03$ $m_0$ and $m^*_{h} \approx 0.27 \pm 0.01$ $m_0$ for electrons and holes at the top conduction band of MoSe$_2$ and top valence band of WSe$_2$, respectively. These values differ significantly from the corresponding reported effective masses. On the other hand, assuming a spin-triplet configuration for IX$^{H}$ results in estimated carrier effective masses that match well previously calculated and experimental values \cite{kormanyos2015k,larentis2018large, goryca2019revealing}, providing confidence that both trapped IX$^{H}$ and IX$^{R}$ originate from optical transitions involving the lowest split conduction band of MoSe$_2$ at $\pm$K.
\end{small}\\

\section{Data availability}
\begin{small}  
Data described in this paper and presented in the Supplementary materials are available online at https://researchportal.hw.ac.uk/en/persons/brian-d-gerardot/datasets/
\end{small}

\end{document}